# Missing-modality Enabled Multi-modal Fusion Architecture for Medical Data

Muyu Wang[1,2], Shiyu Fan[1,2], Yichen Li[1,2], Hui Chen[1,2*]

1. School of Biomedical Engineering, Capital Medical University, No.10, Xitoutiao, You An Men, Fengtai District, Beijing, China, 100069.

2. Beijing Key Laboratory of Fundamental Research on Biomechanics in Clinical Application, Capital Medical University, No.10, Xitoutiao, You An Men, Fengtai District, Beijing, China, 100069.

* Corresponding author

## Abstract

Fusing multi-modal data can improve the performance of deep learning models. However, missing modalities are common for medical data due to patients' specificity, which is detrimental to the performance of multi-modal models in applications. Therefore, it is critical to adapt the models to missing modalities. This study aimed to develop an efficient multi-modal fusion architecture for medical data that was robust to missing modalities and further improved the performance on disease diagnosis.

X-ray chest radiographs for the image modality, radiology reports for the text modality, and structured value data for the tabular data modality were fused in this study. Each modality pair was fused with a Transformer-based bi-modal fusion module, and the three bi-modal fusion modules were then combined into a tri-modal fusion framework. Additionally, multivariate loss functions were introduced into the training process to improve model's robustness to missing modalities in the inference process. Finally, we designed comparison and ablation experiments for validating the effectiveness of the fusion, the robustness to missing modalities and the enhancements from each key component. Experiments were conducted on MIMIC-IV, MIMIC-CXR with the 14-label disease diagnosis task. Areas under the receiver operating characteristic curve (AUROC), the area under the precision-recall curve (AUPRC) were used to evaluate models' performance. The experimental results demonstrated that our proposed multi-modal fusion architecture effectively fused three modalities and showed strong robustness to missing modalities. This method is hopeful to be scaled to more modalities to enhance the clinical practicality of the model.

# 1 Introduction

The volume and variety of medical data has grown rapidly in recent years, which can be used as a source of tremendous amounts of data for the development of deep learning models, laying the foundation of clinical decision support systems and precision medicine[1, 2]. Medical data are presented in a variety of modalities, such as well-organized tabular data (e.g., demographics and laboratory results), free-style texts (e.g., radiology reports and progress notes), images (e.g., X-rays and magnetic resonance imaging [MRI]), signals (e.g., electrocardiogram and electroencephalogram), and videos such as endoscopy. It has been shown that multi-modal data can improve the performance of deep learning models[3]. Deep learning models built on multi-modal medical data can have high diagnostic performance, which can help reduce medical costs and solve the shortage of clinical experts[4-7].

Most medical multi-modal fusion studies have focused on two modalities[8], such as the fusion of chest radiographs and tabular data for cardiomegaly diagnosis[9], the fusion of chest radiographs and radiology reports for multi-label classification of chest diseases[10], the fusion of MRI and tabular data for dementia diagnosis[11], and the fusion of physiological time series and clinical notes for early prediction of sepsis[12]. Few study fused three medical modalities of pathlology images, medical record text and tabular pathology features for disease diagnoses[13]. Although bi-modal fusion (BiMF) models were relatively easy to build, they did not follow the practice of clinicians, who make diagnostic decisions using all possible modalities of patient data.

When more medical modalities are considered for fusion, missing modalities are inevitable in the real-world clinical application scenarios and become a critical issue that is not conducive to model application. Therefore, deep learning methods, such as autoencoder[14] and generative adversarial network (GAN)[15, 16] have been developed for the generation and imputation of the missing modality based on the feature extraction from the original modal data. These methods may not be suitable for all types of medical modalities and require massive sample data for training[17].

In this study, we proposed a Transformer-based tri-modal fusion (TriMF) architecture, including three feature embedding networks for each individual modality and a multi-modal fusion framework with a multivariate loss function. This architecture was adopted for the fusion of chest radiographs, corresponding radiology reports, and tabular data for the diagnosis of thoracic diseases. We aimed to enhance the effect of multi-modal fusion while improving the model's robustness to missing modalities.

# 2 Method

In this section, we introduced the proposed multi-modal fusion architecture, including feature embedding networks for multi-modalities, a TriMF framework and multivariate loss functions. We designed a series of experiments to evaluate the fusion performance by applying the proposed multi-modal fusion models to a 14-label classification task.

## 2.1 Datasets

We used the open access MIMIC-IV and MIMIC-CXR datasets[18-20]. Patients' demographics and laboratory tests were maintained as records in the MIMIC-IV dataset, and the digital chest radiographs, the associated radiology reports and CheXpert labels were maintained in the MIMIC-CXR dataset. Five demographic characteristics (gender, age, source of admission, insurance, and ethnicity) and 46 high-frequency (> 50%) laboratory tests were extracted from the MIMIC-IV dataset to form a tabular modality. Anteroposterior chest radiographs extracted from the MIMIC-CXR dataset were considered as an image modality. Free-text radiology reports for these radiographs were extracted simultaneously. The "Findings" section of a radiology report provides an objective description of imaging features and was therefore selected as the text modality for fusion. In addition, patients (samples) in the dataset were assigned one of the 14 disease labels, which were used in the classification task of the study. We excluded samples without the complete three modalities or disease labels and finally obtained 23,421 samples as our experimental dataset. More details about the demographic characteristics, laboratory tests, and disease labels of the study samples are listed in Tables S1, S2, and S3, respectively.

## 2.2 Multi-modal Fusion Framework

Data from the three individual modalities were first seperately embedded into a low-dimensional real number space by pre-trained models and simple neural networks, which were considered to help avoid over-fitting. Instead of fine-tuning these embedding modules to obtain the definite representations, we trained them together with the multi-modal fusion modules in order to build an end-to-end model (Figure 1).

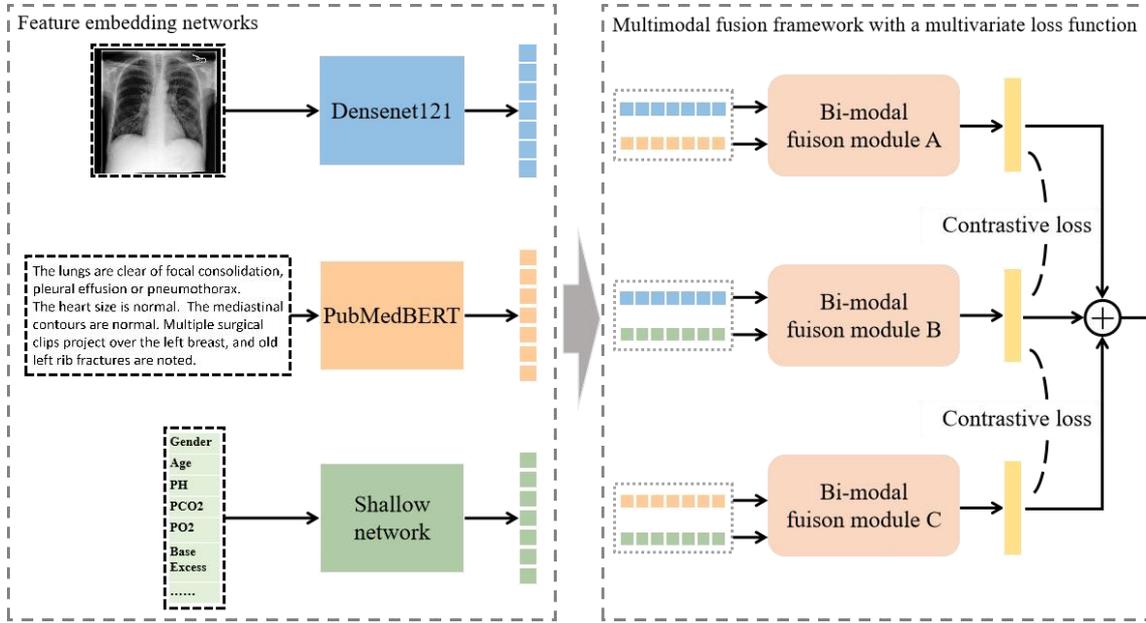

**Fig. 1.** The proposed TriMF architecture for medical data.

**2.2.1 Feature embedding for each modality**

*Feature embedding for images*

Chest radiographs of various size were first resized to a uniform size of 224*224 pixels for ease of embedding. The pre-trained Densenet-121 network[21] was used to extract features from these images. A total of 2048 feature maps of 7*7 dimension were extracted on the layer closest to the last pooling layer. Each feature map was flattened into a 49-dimensional vector, and we finally obtained a 49*2048 dimensional embedding representation of each chest radiograph image.

*Feature embedding for text*

The length of the "Findings" section of all radiology reports ranged from 10 to 280 words (Figure S1). The optimal text length for embedding was determined by trial and error to be 150 words, while the vast majority of "Findings" (23269 of 23421 samples, 99.35%) were less than 150 words. Therefore, each report was truncated to 150 words before being fused. Due to its good performance in embedding medical text, PubMedBERT[22], a well pre-trained model using PubMed as the corpus, was used for text embedding in the current study. Each word was embedded into a 768-dimensional vector, and the final embedding dimension of each report was 150*768.

*Feature embedding for tabular data*

We built a shallow neural network with one input neuron and 256 output neurons to embed each tabular feature into a 256-dimensional vector. When the 5 demographic characteristics and 46 laboratory tests for a patient (sample) were treated as a 51*1 vector, the tabular data were finally encoded as embeddings of 51*256 dimension.

### 2.2.2 The BiMF module

The central idea behind the proposed TriMF framework was first to fuse two modalities each by a BiMF module and then to fuse the three BiMF representations together. The BiMF module contained a stacked fusion block and a low-rank multi-modal fusion structure, as shown in Figure 2.

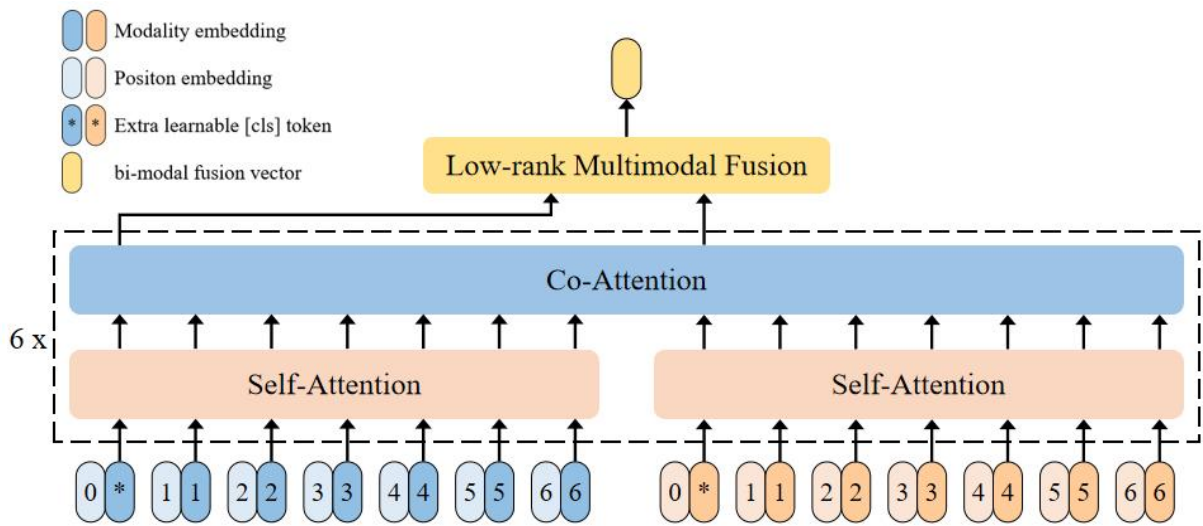

**Fig. 2.** The BiMF module based on self-attention (SA), co-attention (CA), and low-rank multi-modal fusion (LMF) components. The combination of SA and CA components was responsible for the information interaction between two modalities. They were stacked for six layers alternately in this study. The LMF component fused two [cls] tokens from the BiMF encoder into a fusion vector of arbitrary dimension.

The BiMF module contained two types of multi-head attention units, the self-attention (SA) unit and the co-attention (CA) unit. The SA unit consisted of a feed-forward layer with two fully-connected layers with GeLU activation and a multi-head attention layer[23]. Taking the embedding vector of a modality as input, the multi-head attention layer could learn the relationship between tokens within a modality. Furthermore, residual connection and layer normalization were applied to facilitate optimization. The CA unit was composed of two symmetric multi-head attention layers, which took the embedding vectors of two modalities as input, one as queries and the other as keys and values. The symmetric structure helped to learn the pairwise relationship between two modalities. The fusion block, which was the combination of two SA units and one CA unit, was stacked for six layers, resulting in a deep BiMF encoder.

For the encoder input, we further added learnable position embeddings and inserted an extra learnable classification token ([cls] token) at the beginning of the embedding sequence of each modality. So far, we had gotten two fusion features from one BiMF encoder, where the two final hidden vectors of [cls] tokens represented the modalities that would be fused by an LMF structure[24]. LMF was an effective fusion method improved from the tensor fusion network[25] by decomposing the weights in the tensor fusion model into low-rank factors. This fusion vector of modalities A and B was calculated as:

$$h = \left(\sum_{i=1}^{r} w_A^{(i)} \cdot z_A\right) \circ \left(\sum_{i=1}^{r} w_B^{(i)} \cdot z_B\right) \tag{1}$$

where $r$ (here 128) was the rank of the decomposition tensor. $\{w_a^{(i)}, w_b^{(i)}\}_{i=1}^{r}$ were the corresponding low-rank factors of modalities A and B, and $z_A$ and $z_B$ were the final hidden vectors of [cls] tokens from the two modalities, respectively. The dimension of each fusion vector from the above BiMF module could be set to be either the same or different for each modality pair. In this study, the fusion vectors were set to have the same dimension of 256.

The final 256-dimensional TriMF vector was the simple sum of the three BiMF vectors. The advantage of fusing any modality pair into a BiMF vector of the same dimension as that of the final TriMF vector was that the TriMF framework could still work even with a missing modality. At this moment, two out of the three BiMF modules would stop work, and the remaining one would provide the output of the TriMF model.

## 2.3 Model training

The training task was a 14-label classification descibed in section 2.1. The classifier was a linear layer with 256 input neurons for the 256-deimensional TriMF vector and 14 output neurons for 14 labels.

### 2.3.1 Loss function for the training process

Although the proposed TriMF model could deal with the imcompleteness of modalities in the following inference process, the fusion representation and further the classification performance will definitely be affected by the missing modality. For example, the BiMF representation of image and text would be different from the TriMF representation of image, text and tabular data, resulting in the different classification results. Therefore, we adopted a fusion representation contrastive loss (FRCL) function mechanism in the training process to improve the similarity between TriMF and BiMF representations to minimize the negative effect of the missing modality on the classification performance.

The contrastive loss function was in the form of mean square error:

$$L_{FRCL}(F_1, F_2) = \sum_{i=1}^{N}(F_{1i} - F_{2i})^2 \qquad (2)$$

where $F_1$ and $F_2$ were the fusion vectors from two BiMF modules, and $N=256$ was the dimension of these fusion representations. The final loss function being used to optimize the framework's parameters was described as:

$$L = \lambda_1 L_{clf} + \lambda_2 L_{FRC}(F_{I,T}, F_{I,S}) + \lambda_3 L_{FRC}(F_{I,T}, F_{T,S}) + \lambda_4 L_{FRC}(F_{I,S}, F_{T,S}) \qquad (3)$$

where $F_{I,T}$, $F_{I,S}$, and $F_{T,S}$ were the BiMF vectors of modalities image and text, image and tabular, and text and tabular, respectively. $L_{clf}$ was the categorical loss in the form of binary cross-entropy. The weights $\lambda_1$, $\lambda_2$, $\lambda_3$, and $\lambda_4$ were set by trial and error to 1, 3, 3, and 3, respectively. Thus, BiMF representations were encouraged to be as close as possible to the TriMF representation in the embedding space, which could promote the robustness of the model to missing-modal data.

### 2.3.2 Training details

We used an Adam optimizer, with a weight decay of 5e-4 and a batch size of 16 at a maximum of 100 epochs of parameter optimization. The learning rate started at 0.001 and decayed at an exponential rate of 0.8 when the validation loss has stopped decreasing for 2 epochs. Training was stopped when there was no improvement in validation loss for six consecutive epochs. For each model, we selected the optimal parameter set that produced the least validation loss. All models were implemented in Pytorch 1.10 and trained on a workstation equipped with an Intel Xeon Gold 5218, 512 GB RAM, and a 16G NVIDIA Tesla T4 GPU.

## 2.4 Model Evaluation

The proposed multi-modal fusion framework was evaluated on a 14-label classification task. Two common performance metrics, area under the receiver operating characteristic curve (AUROC) and area under the precision-recall curve (AUPRC), were used for each of the 14 labels, and the average AUROC and AUPRC were used for the overall performance evaluation. The training, validation, and test sets were randomly divided at a ratio of 8:1:1.

### 2.4.1 Comparative Experiments

The comparative experiments were conducted to compare the proposed TriMF framework with other models using the same dataset, and with other models using one or two modalities to show the advantages of the fusion of three modalities.

The proposed model was first compared with MedViLL[26], a Transformer-based multi-modal fusion model. The MedViLL model was originally pre-trained on the MIMIC-CXR dataset and then used to fuse chest radiographs and reports containing both "Findings" and "Impression" sections. In this study, radiology reports containing only the "Findings" section were fused with chest radiographs. Therefore, in the current experiment, the MedViLL model was first fine-tuned and then tested using the chest radiographs and reports in our sample set to ensure that the dataset was the same as that used for the proposed model.

The proposed multi-modal fusion model was then compared with the PubMedBERT model trained and tested with text modality only, ResNet-50 with image modality only, shallow neural network with tabular data modality only, and the proposed BiMF models with two modalities of image and text, image and tabular data, and text and tabular data, respectively.

### 2.4.2 Experiments on the robustness of the model

One of the three modalities of the test set was separately masked out to construct three incomplete modality test sets. For each incomplete modality test set, the proposed multi-modal fusion framework was trained on the training set containing A) all three modalities (the TriMF framework was used), and B) two modalities as in the incomplete modality test set (the BiMF module was used), respectively. Both were tested on the incomplete modality test set. The classification performance of Model A was compared to that of Model B to evaluate the performance improvement of the model trained with the complete versus incomplete modality training sets when tested on the incomplete modality test set. To evaluate the model's robustness to missing modalities, the classification performance of Model A (trained on the complete and tested on the incomplete modality samples) was also compared to that of the model trained and tested all on the complete modality samples.

**2.4.3 Ablation Experiments**

In the proposed multi-modal fusion architecture, we introduced SA units and the LMF mechanism in the BiMF module, and the FRCL function in the training process, which were considered critical to the architecture. To identify the effect of these key components on the classification performance, we conducted ablation experiments focusing on SA units, LMF, and contrastive loss function, respectively. When SA units were ablated, they were replaced by CA units. When the LMF was ablated, all output [cls] tokens from the BiMF modules were simply concatenated and fed into a classifier. When the multi-modal architecture was not trained under the proposed contrastive loss function, the binary cross-entropy for classification was used as a substitute. Samples with all three modalities were used in these ablation experiments.

# 3 Result

## 3.1 Comparative Experiments

When trained and tested on the same training and test sample sets, the proposed TriMF model outperformed the MedViLL model, regardless of whether AUROC or AUPRC was used in the performance comparisons. The performance of the proposed TriMF model was higher than that of MedViLL for 11 and 9 out of 14 labels, respectively, which contributed a higher average performance (AUROC 0.914 vs. 0.870 and AUPRC 0.552 vs. 0.484) to the proposed model (Figure 3 and Table S4).

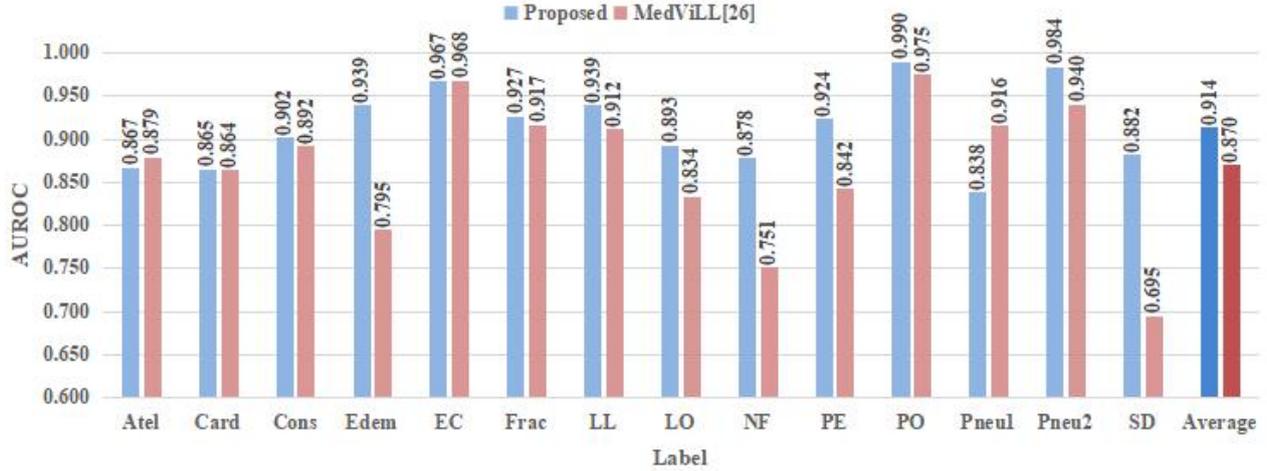

Fig. 3 Performance comparison between the proposed TriMF model and the MedViLL model[26] in a classification task for 14 lung related diseases. Meanings of the 14 labels are listed in Table S1.

The performance of classification models based on different single modalities or combinations of modalities varied. The more modalities involved, the better the model, as shown in Tables 1 and S5. The proposed TriMF model trained and tested on all three modalities (i.e., image, text, and tabular data) achieved a higher performance than any of those trained and tested on one or two modalities on average (AUROC 0.914 and AUPRC 0.552) as well as for 10 out of 14 labels. When two modalities were fused using the proposed BiMF module, the bi-modal combination of the image and text (AUROC 0.879) outperformed both the single image and single text modalities (0.732 and 0.835, respectively). This was also the case for the bi-modal combinations of image and tabular data (AUROC 0.815 vs. 0.732 and 0.568) and text and tabular data (0.875 vs. 0.835 and 0.568). It is worth mentioning that the BiMF module trained and tested on the modality combination of image and text showed a higher performance (AUROC 0.879 and AUPRC 0.502) than MedViLL (0.870 and 0.484) with the same data set used.

In addition, the text modality tended to play the most important role among the three modalities. Among the three unimodal-based models, the model based on the text modality alone achieved the best performance on eight labels, and the model based on the image modality alone performed best on the remaining six labels (average AUROC, 0.869 and 0.819, respectively). For the three bi-modal based models, the models based on the bi-modal combination of image and text, image and tabular data, and text and tabular data showed the best performance on eight, two, and four labels, with an average AUROCs of 0.882, 0.902, and 0.930, respectively. However, the model based on the fusion of image and text outperformed the TriMF model on three labels, indicating a certain negative effect of tabular data on the fusion of three modalities.

Table 1 Performance comparison among different modality combinations used in the proposed TriMF model in a classification task for 14 lung related diseases.

| AUROC | Single modality | | | Two modalities | | | All three modalities |
|---|---|---|---|---|---|---|---|
| | Image only | Text only | Tabular data only | Image and text | Image and tabular data | Text and tabular data | |
| Atel | 0.764 | 0.729 | 0.548 | **0.882** | 0.712 | 0.865 | 0.867 |
| Card | 0.652 | 0.742 | 0.531 | 0.860 | 0.714 | 0.829 | **0.865** |
| Cons | 0.698 | 0.845 | 0.589 | 0.881 | 0.821 | 0.860 | **0.902** |
| Edem | 0.882 | 0.856 | 0.599 | 0.901 | 0.845 | 0.925 | **0.939** |
| EC | 0.852 | 0.889 | 0.502 | 0.869 | 0.845 | 0.890 | **0.967** |
| Frac | 0.507 | 0.926 | 0.567 | 0.807 | 0.899 | 0.876 | **0.927** |
| LL | 0.691 | 0.917 | 0.605 | 0.879 | 0.886 | 0.920 | **0.939** |
| LO | 0.839 | 0.788 | 0.542 | 0.891 | 0.786 | 0.884 | **0.893** |
| NF | 0.835 | 0.825 | 0.547 | **0.894** | 0.817 | 0.871 | 0.878 |
| PE | 0.855 | 0.819 | 0.553 | 0.918 | 0.787 | 0.909 | **0.924** |
| PO | 0.307 | 0.932 | 0.664 | 0.832 | 0.905 | 0.723 | **0.990** |
| Pneu1 | 0.737 | 0.718 | 0.572 | 0.833 | 0.702 | 0.825 | **0.838** |
| Pneu2 | 0.881 | 0.938 | 0.574 | 0.965 | 0.931 | **0.986** | 0.984 |
| SD | 0.743 | 0.763 | 0.562 | **0.900** | 0.759 | 0.888 | 0.882 |
| Average | 0.732 | 0.835 | 0.568 | 0.879 | 0.815 | 0.875 | **0.914** |

Note: The largest values are bolded for each label. Meanings of the 14 labels are listed in Table S1. AUROC: area under the receiver operating characteristic curve.

## 3.2 The robustness of the TriMF framework

Tables 2 and S6 present the performance of our TriMF model to classify the 14 labels when one modality was missing. After the proposed classification model was built on training samples with the complete modalities, the model showed only a slight degradation in classification accuracy when tested on samples with one missing modality, compared to when tested on samples with the complete modalities. The average AUROCs decreased by 0.002, 0.071, and 0.003 from 0.914 when tabular data, text, and image were missing, respectively. The result suggests that our architecture successfully promoted the robustness to missing modalities. In addition, the classification accuracy decreased the most in the absence of the text modality and the least in the absence of the tabular data modality, indicating that the tabular data modality had the least impact on the model's robustness, while text had the most.

On the other hand, when the classification task was performed with two modalities, the classification model built on training samples containing all three modalities outperformed those built on training samples with the same two modalities as in the test set on average and for most of the labels. For example, our TriMF model built on the image, text, and tabular data modalities outperformed the BiMF model built on the image and text modalities when tested on the test samples without the tabular data modality on average (AUROC 0.912 vs. 0.879, and AUPRC 0.540 vs. 0.502) and for 10 out of 14 labels. This indicates that even if a specific modality was not used in the inference process, the TriMF model was still able to effectively fuse the modality with other modalities in the training process, thus improving the classification performance of the model. Taken together, these results provide important insights into that our proposed architecture not only promotes the robustness to missing modalities, but also enhances the multi-modal fusion effect.

Table 2 Performance (in AUROC) of the classification models tested on incomplete modality sets.

| Labels | Test modality: Im_Tx | | Test modality: Im_Ta | | Test modality: Tx_Ta | |
|---|---|---|---|---|---|---|
| | Im_Tx_Ta | Im_Tx | Im_Tx_Ta | Im_Ta | Im_Tx_Ta | Tx_Ta |
| Atel | 0.866 | 0.882 | 0.730 | 0.712 | 0.861 | 0.865 |
| Card | 0.865 | 0.860 | 0.788 | 0.714 | 0.862 | 0.829 |
| Cons | 0.902 | 0.881 | 0.861 | 0.821 | 0.906 | 0.860 |
| Edem | 0.937 | 0.901 | 0.873 | 0.845 | 0.936 | 0.925 |
| EC | 0.966 | 0.869 | 0.840 | 0.845 | 0.960 | 0.890 |
| Frac | 0.926 | 0.807 | 0.896 | 0.899 | 0.925 | 0.876 |
| LL | 0.921 | 0.879 | 0.926 | 0.886 | 0.935 | 0.920 |
| LO | 0.890 | 0.891 | 0.787 | 0.786 | 0.891 | 0.884 |
| NF | 0.876 | 0.894 | 0.850 | 0.817 | 0.882 | 0.871 |
| PE | 0.923 | 0.918 | 0.766 | 0.787 | 0.918 | 0.909 |
| PO | 0.990 | 0.832 | 0.940 | 0.905 | 0.986 | 0.723 |
| Pneu1 | 0.834 | 0.833 | 0.775 | 0.702 | 0.831 | 0.825 |
| Pneu2 | 0.984 | 0.965 | 0.976 | 0.931 | 0.987 | 0.986 |
| SD | 0.882 | 0.900 | 0.794 | 0.759 | 0.877 | 0.888 |

| | Average | 0.912 | 0.879 | 0.843 | 0.815 | 0.911 | 0.875 |

Note: The modalities used in the training process are shown in the second row of the column heading. Meanings of the 14 labels are listed in Table S1. Im_Tx_Ta: modality combination of all three modalities; Im_Tx: modality combination of image and text; Im_Ta: modality combination of image and tabular data; Tx_Ta: modality combination of text and tabular data.

## 3.3 Ablation Experiments

The results of the ablation study are reported in Tables 3 and S7. Due to the absence of SA, LMF, and FRCL function, the average AUROC of our model decreased by 0.030, 0.029, 0.034, and the average AUPRC decreased by 0.053, 0.056, 0.060, respectively. It could be seen that all three components improved the overall performance of the model. In particular, the complete model outperformed all three ablation models in terms of both AUROC and AUPRC, on four labels (enlarge cardiomediastinum [EC], fracture [Frac], lung lesion [LL], and pleural other [PO]) with the lowest positive rates (3.2%, 0.9%, 1.6%, and 0.4%, respectively). Taken together, all three components improved the average performance, but mainly improved the classification of extremely unbalanced labels and slightly sacrificed the classification of other labels. Specifically, ablation of SA led to an increase in AUROC and a decrease in AUPRC for two labels (cardiomegaly [Card] and pleural effusion [PE]), similar to ablation of LMF for three labels (lung opacity [LO], no finding [NF], and support devices [SD]) and ablation of FRCL function for two labels (lung opacity [LO] and no finding [NF]). This implied that the addition of any of these components would boost the true positive rate of the classification task for some labels, which is beneficial for a disease diagnostic model.

Table 3 Model performance in the ablation experiments around SA, LMF and FRCL components.

| | Labels | Atel | Card | Cons | Edem | EC | Frac | LL | LO | NF | PE | PO | Pneu1 | Pneu2 | SD | Mean |
|---|---|---|---|---|---|---|---|---|---|---|---|---|---|---|---|---|
| AUROC | Proposed | 0.867 | 0.865 | 0.902 | 0.939 | 0.967 | 0.927 | 0.939 | 0.893 | 0.878 | 0.924 | 0.990 | 0.838 | 0.984 | 0.882 | 0.914 |
| | w/o SA | 0.878 | 0.869 | 0.878 | 0.938 | 0.895 | 0.819 | 0.899 | 0.902 | 0.894 | 0.932 | 0.756 | 0.832 | 0.981 | 0.897 | 0.884 |
| | w/o LMF | 0.875 | 0.831 | 0.863 | 0.937 | 0.890 | 0.878 | 0.916 | 0.895 | 0.887 | 0.931 | 0.792 | 0.825 | 0.986 | 0.888 | 0.885 |
| | w/o FRCL | 0.875 | 0.829 | 0.860 | 0.935 | 0.890 | 0.876 | 0.920 | 0.894 | 0.891 | 0.929 | 0.723 | 0.825 | 0.986 | 0.888 | 0.880 |
| AUPRC | Proposed | 0.541 | 0.319 | 0.314 | 0.793 | 0.479 | 0.221 | 0.415 | 0.701 | 0.807 | 0.794 | 0.444 | 0.328 | 0.828 | 0.751 | 0.552 |
| | w/o SA | 0.570 | 0.298 | 0.367 | 0.802 | 0.391 | 0.030 | 0.111 | 0.732 | 0.828 | 0.803 | 0.036 | 0.360 | 0.883 | 0.769 | 0.499 |

| | | | | | | | | | | | | | | |
|---|---|---|---|---|---|---|---|---|---|---|---|---|---|---|
| w/o LMF | 0.578 | 0.243 | 0.273 | 0.807 | 0.363 | 0.087 | 0.274 | 0.697 | 0.812 | 0.810 | 0.026 | 0.335 | 0.884 | 0.749 | 0.496 |
| w/o FRCL | 0.575 | 0.243 | 0.275 | 0.807 | 0.367 | 0.070 | 0.256 | 0.693 | 0.822 | 0.810 | 0.015 | 0.329 | 0.879 | 0.752 | 0.492 |

Notes: Meanings of the 14 labels are listed in Table S1. AUROC: area under the receiver operating characteristic curve, AUPRC: area under the precision-recall curve, w/o: without.

## 4 Discussion

Multi-modal medical data fusion model based on deep learning has achieved remarkable results. Many researchers have attempted to improve the performance of clinical tasks by fusing multi-modal data, including images, free-text reports, clinical audio, biological signals, laboratory tests, and so on. Similar to studies in the general domain, multi-modal fusion studies in the medical domain have mostly focused on bi-modal fusion of the image and text modalities[9-12]. However, many types of modalities are generated in clinical practice [27], which may facilitate the use of deep learning models to learn better patient representations[3]. Therefore, we designed a delicate multi-modal fusion architecture with the goal of effectively fusing not only the image and text modality, but also the structured data modality. Through the multiple perspectives validation experiments, showed that the proposed multi-modal fusion architecture effectively fused the three modalities and improved the classification performance of models based on the fused information.

Missing modality is a common problem for medical data, which can lead to a dramatic performance degradation of multi-modal fusion models in real-world clinical application[28, 29]. To solve this problem, the generation of missing modalities based on other modalities using e.g. autoencoder[14] and GAN[16] is the prevalent idea of current research. However, these methods have constraints on the type of modality and require a large number of samples. In this paper, instead of generating the missing modality, we concentrated on constructing a model that can accept the missing modality data as input by combining several separate BiMF modules. To reduce the negative impact of missing modalities on the classification performance, we proposed a multivariate loss function containing FRCL and classification loss in the training of the multi-modal fusion model to make the fused patient representation with and without missing modalities as similar as possible. The subtle design of the BiMF module and the multivariate loss function improved the robustness of the multi-modal fusion framework to missing-modal data, as well as its scalability, allowing us to fuse more modalities by adding BiMF.

Feature concatenation, autoencoder, attention mechanism, etc. were previously the most prevalent multi-modal deep learning methods on medical data[30-32]. With the success of Transformer-based model on natural texts and images[33, 34], Transformer block has been used in multi-modal fusion[35-37]. The current Transformer-based medical multi-modal fusion models were mostly based on SA[10, 26]. According to Li et al.[38], SA and CA backbones were good at aligning low-level and high-level semantics, respectively. Yu et al.[36] proved that alternate stacking of SA and CA could improve the performance of Transformer-based multi-modal models, which was also confirmed by our study. When SA units were incorporated into the dual-stream architecture based on CA, the average AUROC and AUPRC increased from 0.884 to 0.914, and 0.499 to 0.552, respectively. In particular, MedViLL is an SA-based BiMF model pre-trained on MIMIC-CXR, which performed worse than not only the proposed TriMF model (AUROC 0.870 vs. 0.914), but also our BiMF module (AUC 0.870 vs. 0.879). These indicate that the Transformer-based fusion framework combining SA and CA can effectively fuse two or even three modalities on medical data.

There were still some limitations in this study. First, although this architecture could fuse more modalities by using more BiMF modules, this would lead to a massive increase in the number of parameters. Second, the modalities included in the current study were still limited. In addition, there are various forms of submodalities for the medical modalities, such as X-ray, CT, and MRI for the image modality, and radiology reports, pathology reports, and hospital admission notes for the text modality. More modalities and more submodalities would introduce a more severe modal missing problem. Whether the proposed architecture is still robust to missing more modalities needs to be verified. Finally, this study only conducted experiments on an English public dataset. The proposed architecture should be validated and evaluated on a Chinese dataset, and hopefully applied to a real clinical scenario in China.

## 5 Conclusions

In this study, we proposed a multi-modal fusion architecture based on Transformer. This architecture could effectively fuse three medical modalities and improve the diagnosis performance, while is robust to modal-incomplete data. This study provided a novel idea for dealing with missing modalities in multi-modal medical data fusion. It has the potential to be scaled to more modalities with the enhanced clinical practicality.

## Acknowledgments

This work was supported by the National Natural Science Foundation of China (grant number 81971707) and the Beijing Natural Science Foundation (grant number L222006).

## Authors' Contributions

Hui Chen: Conceptualization, Writing- Reviewing and Editing. Muyu Wang: Data curation, Methodology, Writing- Original draft preparation. Shiyu Fan: Data curation. Yichen Li: Data curation.

## Conflicts of Interest

None declared.

## Abbreviations

AUPRC: area under the precision-recall curve

AUROC: area under the receiver operating characteristic curve

BiMF: bi-modal fusion

CA: co-attention

FRCL: fusion representation contrastive loss

GAN: generative adversarial network

LMF: Low-rank Multi-modal Fusion

MIMIC: Medical Information Mart for Intensive Care

MRI: magnetic resonance imaging

SA: self-attention

TriMF: tri-modal fusion

# References

[1] R.T. Sutton, D. Pincock, D.C. Baumgart, D.C. Sadowski, R.N. Fedorak and K.I. Kroeker, "An overview of clinical decision support systems: benefits, risks, and strategies for success," npj Digit. Med., vol. 3, no. 1, pp. 17, 2020, doi: 10.1038/s41746-020-0221-y.

[2] E.J. Topol, "High-performance medicine: the convergence of human and artificial intelligence," Nat. Med., vol. 25, no. 1, pp. 44-56, 2019, doi: 10.1038/s41591-018-0300-7.

[3] Y. Huang, C. Du, Z. Xue, X. Chen, H. Zhao and L. Huang, "What Makes Multi-modal Learning Better than Single (Provably)," arXiv:2106.04538, 2021.

[4] C. MaoL. Yao and Y. Luo, "ImageGCN: Multi-Relational Image Graph Convolutional Networks for Disease Identification with Chest X-rays," IEEE Trans. Med. Imaging, vol. 8, no. 41, pp. 1990-2003, 2022, doi: 10.1109/TMI.2022.3153322.

[5] U. Kamal, M. Zunaed, N.B. Nizam and T. Hasan, "Anatomy-XNet: An Anatomy Aware Convolutional Neural Network for Thoracic Disease Classification in Chest X-Rays," IEEE J. Biomed. Health Inform., vol. 26, no. 11, pp. 5518-5528, 2022, doi: 10.1109/JBHI.2022.3199594.

[6] A. Casey et al., "A systematic review of natural language processing applied to radiology reports," BMC Med. Inform. Decis. Mak., vol. 21, no. 1, pp. 179, 2021, doi: 10.1186/s12911-021-01533-7.

[7] S. Jeon, Z. Colburn, J. Sakai, L. Hung and K.Y. Yeung, "Application of Natural Language Processing and Machine Learning to Radiology Reports," in Proc. 12th ACM Conf. Bioinf. Comput. Biol. Health Informat., pp. 1-9, 2021.

[8] A. Kline et al., "Multimodal machine learning in precision health: A scoping review," npj Digit. Med., vol. 5, no. 1, pp. 171, 2022, doi: 10.1038/s41746-022-00712-8.

[9] D. Grant, B.W. Papież, G. Parsons, L. Tarassenko and A. Mahdi, "Deep Learning Classification of Cardiomegaly Using Combined Imaging and Non-imaging ICU Data," in Machine Learning in Medical Imaging, Springer, pp. 547-558, 2021.

[10] G. JacenkówA.Q. O'Neil and S.A. Tsaftaris, "Indication as Prior Knowledge for Multimodal Disease Classification in Chest Radiographs with Transformers," in 2022 IEEE 19th

International Symposium on Biomedical Imaging (ISBI), IEEE, pp. 1-5, 2022.

[11] S. Qiu et al., "Multimodal deep learning for Alzheimer's disease dementia assessment," Nat. Commun., vol. 13, no. 1, pp. 3404, 2022, doi: 10.1038/s41467-022-31037-5.

[12] Y. Wang, Y. Zhao, R. Callcut and P. Linda, "Integrating Physiological Time Series and Clinical Notes with Transformer for Early Prediction of Sepsis," arXiv:2203.14469, 2022.

[13] M. Song, X. Shi, Y. Zhang and B. Li, "Multimodal Breast Cancer Diagnosis Based on Multi-level Fusion Network," in ISAIR 2022: Artificial Intelligence and Robotics, pp. 224-239, 2022.

[14] Y. Xu et al., "Explainable Dynamic Multimodal Variational Autoencoder for the Prediction of Patients with Suspected Central Precocious Puberty," IEEE J. Biomed. Health Inform., vol. 26, no. 3, pp. 1362-1373, 2021, doi: 10.1109/JBHI.2021.3103271.

[15] J. YoonJ. Jordon and M. van der Schaar, "GAIN: Missing Data Imputation using Generative Adversarial Nets," in International Conference on Machine Learning, PMLR, pp. 5689-5698, 2018.

[16] T. Zhou, S. Canu, P. Vera and S. Ruan, "Feature-enhanced generation and multi-modality fusion based deep neural network for brain tumor segmentation with missing MR modalities," Neurocomputing, vol. 466, pp. 102-112, 2021, doi: 10.1016/j.neucom.2021.09.032.

[17] Y. Liu, H. Ishibuchi, G.G. Yen, Y. Nojima and N. Masuyama, "Handling Imbalance Between Convergence and Diversity in the Decision Space in Evolutionary Multi-Modal Multi-Objective Optimization," IEEE Trans. Evol. Comput., vol. 24, no. 3, pp. 551-565, 2020, doi: 10.1109/TEVC.2019.2938557.

[18] A.E.W. Johnson et al., "MIMIC-CXR, a de-identified publicly available database of chest radiographs with free-text reports," Sci. Data, vol. 6, pp. 317, 2019, doi: 10.1038/s41597-019-0322-0.

[19] A.E. Johnson et al., "MIMIC-CXR-JPG, a large publicly available database of labeled chest radiographs," arXiv:1901.07042, 2019.

[20] A. Johnson, L. Bulgarelli, T. Pollard, S. Horng, L.A. Celi and R. Mark, "Mimic-iv (version 1.0)," PhysioNet, 2020.

[21] G. Huang, Z. Liu, L. van der Maaten and K.Q. Weinberger, "Densely Connected Convolutional Networks," in Proc. IEEE Conf. Comput. Vis. Pattern Recognit., IEEE, pp. 4700-4708, 2017.

[22] Y. Gu et al., "Domain-Specific Language Model Pretraining for Biomedical Natural Language Processing," ACM Transactions on Computing for Healthcare, vol. 3, no. 1, pp. 1-23, 2022, doi: 10.1145/3458754.

[23] A. Vaswani et al., "Attention Is All You Need," in Adv. Neural Inf. Process. Syst., pp. 5998-6008, 2017.

[24] Z. Liu, Y. Shen, V.B. Lakshminarasimhan, P.P. Liang, A. Zadeh and L. Morency, "Efficient Low-rank Multimodal Fusion with Modality-Specific Factors," arXiv:1806.00064, 2018.

[25] A. Zadeh, M. Chen, S. Poria, E. Cambria and L. Morency, "Tensor Fusion Network for Multimodal Sentiment Analysis," arXiv:1707.07250, 2017.

[26] J.H. Moon, H. Lee, W. Shin and E. Choi, "Multi-modal Understanding and Generation for Medical Images and Text via Vision-Language Pre-Training," IEEE J. Biomed. Health Inform., vol. 26, no. 12, pp. 6070-6080, 2021.

[27] A. HaqueA. Milstein and L. Fei-Fei, "Illuminating the dark spaces of healthcare with ambient intelligence," Nature, vol. 585, no. 7824, pp. 193-202, 2020, doi: 10.1038/s41586-020-2669-y.

[28] Q. Suo, Z. Weida, M. Fenglong, Y. Ye, G. Jing and A. Zhang, "Metric Learning on Healthcare Data with Incomplete Modalities," in Int. Joint Conf. Artif. Intell., pp. 3534-3540, 2019.

[29] C. Zhang et al., "M3Care: Learning with Missing Modalities in Multimodal Healthcare Data," in Proceedings of the ACM international conference on knowledge discovery and data mining, pp. 2418-2428, 2022.

[30] W. Ning et al., "Open resource of clinical data from patients with pneumonia for the prediction of COVID-19 outcomes via deep learning," Nat. Biomed. Eng., vol. 4, no. 12, pp. 1197-1207, 2020, doi: 10.1038/s41551-020-00633-5.

[31] T. van Sonsbeek and M. Worring, "Towards Automated Diagnosis with Attentive Multi-modal Learning Using Electronic Health Records and Chest X-Rays," in ML-CDS and CLIP (MICCAI), Springer, pp. 106-114, 2020.

[32] V. Singh, N.K. Verma, Z. Ul Islam and Y. Cui, "Feature Learning Using Stacked Autoencoder

for Shared and Multimodal Fusion of Medical Images," in Computational Intelligence: Theories Applications and Future Directions, Springer, vol. 1, pp. 53-66, 2019.

[33] A. Dosovitskiy et al., "An Image is Worth 16x16 Words: Transformers for Image Recognition at Scale," in Proc. Int. Conf. Learn. Represent., 2021.

[34] J. Devlin, M. Chang, K. Lee and T. Kristina, "BERT: Pre-training of Deep Bidirectional Transformers for Language Understanding," arXiv:1810.04805, 2018.

[35] W. KimB. Son and I. Kim, "ViLT: Vision-and-Language Transformer Without Convolution or Region Supervision," in Int. Conf. Mach. Learn. (ICML), pp. 5583-5594, 2021.

[36] Z. Yu, J. Yu, Y. Cui, D. Tao and Q. Tian, "Deep Modular Co-Attention Networks for Visual Question Answering," in Proc. IEEE Conf. Comput. Vis. Pattern Recognit., pp. 6281-6290, 2019.

[37] Y.H. Tsai, S. Bai, L.P. Pu, J.Z. Kolter, L.P. Morency and R. Salakhutdinov, "Multimodal Transformer for Unaligned Multimodal Language Sequences," in Proc. Conf. Assoc. Comput. Linguistics, vol. 2019, pp. 6558-6569, 2019.

[38] C. Li et al., "SemVLP: Vision-Language Pre-training by Aligning Semantics at Multiple Levels," arXiv:2103.07829, 2022.

校对报告

当前使用的样式是 [IEEE Access New]
当前文档包含的题录共 49 条
有 0 条题录存在必填字段内容缺失的问题
所有题录的数据正常

# References:

[1] R.T. Sutton, D. Pincock, D.C. Baumgart, D.C. Sadowski, R.N. Fedorak and K.I. Kroeker, "An overview of clinical decision support systems: benefits, risks, and strategies for success," *NPJ Digit. Med.*, vol. 3, no. 1, pp. 17, 2020, doi: 10.1038/s41746-020-0221-y.

[2] E.J. Topol, "High-performance medicine: the convergence of human and artificial intelligence," *Nat. Med.*, vol. 25, no. 1, pp. 44-56, 2019, doi: 10.1038/s41591-018-0300-7.

[3] Y. Huang, C. Du, Z. Xue, X. Chen, H. Zhao and L. Huang, "What Makes Multi-modal Learning Better than Single (Provably)," 2021, *arXiv:2106.04538*.

[4] C. MaoL. Yao and Y. Luo, "ImageGCN: Multi-Relational Image Graph Convolutional Networks for Disease Identification with Chest X-rays," *IEEE Trans. Med. Imaging*, vol. 8, no. 41, pp. 1990-2003, doi:

10.1109/TMI.2022.3153322.

[5] U. Kamal, M. Zunaed, N.B. Nizam and T. Hasan, "Anatomy-XNet: An Anatomy Aware Convolutional Neural Network for Thoracic Disease Classification in Chest X-Rays," *IEEE J. Biomed. Health Inform.*, vol. 26, no. 11, pp. 5518-5528, 2022, doi: 10.1109/JBHI.2022.3199594.

[6] A. Casey et al., "A systematic review of natural language processing applied to radiology reports," *BMC Med. Inform. Decis. Mak.*, vol. 21, no. 1, pp. 179, 2021, doi: 10.1186/s12911-021-01533-7.

[7] S. Jeon, Z. Colburn, J. Sakai, L. Hung and K.Y. Yeung, "Application of Natural Language Processing and Machine Learning to Radiology Reports," in *Proc. 12th ACM Conf. Bioinf. Comput. Biol. Health Informat.*, pp. 1-9, 2021.

[8] A. Kline et al., "Multimodal machine learning in precision health: A scoping review," *NPJ Digit. Med.*, vol. 5, no. 1, pp. 171, 2022, doi: 10.1038/s41746-022-00712-8.

[9] D. Grant, B.W. Papież, G. Parsons, L. Tarassenko and A. Mahdi, "Deep Learning Classification of Cardiomegaly Using Combined Imaging and Non-imaging ICU Data," in *Machine Learning in Medical Imaging*, Springer, pp. 547-558, 2021.

[10] G. Jacenków A.Q. O'Neil and S.A. Tsaftaris, "Indication as Prior Knowledge for Multimodal Disease Classification in Chest Radiographs with Transformers," in *2022 IEEE 19th International Symposium on Biomedical Imaging (ISBI)*, IEEE, pp. 1-5, 2022.

[11] S. Qiu et al., "Multimodal deep learning for Alzheimer's disease dementia assessment," *Nat. Commun.*, vol. 13, no. 1, pp. 3404, 2022, doi: 10.1038/s41467-022-31037-5.

[12] Y. Wang, Y. Zhao, R. Callcut and P. Linda, "Integrating Physiological Time Series and Clinical Notes with Transformer for Early Prediction of Sepsis," 2022, *arXiv:2203.14469*.

[13] M. Song, X. Shi, Y. Zhang and B. Li, "Multimodal Breast Cancer Diagnosis Based on Multi-level Fusion Network," in *ISAIR 2022: Artificial Intelligence and Robotics*, pp. 224-239, 2022.

[14] Y. Xu et al., "Explainable Dynamic Multimodal Variational Autoencoder for the Prediction of Patients with Suspected Central Precocious Puberty," *IEEE J. Biomed. Health Inform.*, vol. 26, no. 3, pp. 1362-1373, 2021, doi: 10.1109/JBHI.2021.3103271.

[15] J. Yoon J. Jordon and M. van der Schaar, "GAIN: Missing Data Imputation using Generative Adversarial Nets," in *International Conference on Machine Learning*, PMLR, pp. 5689-5698, 2018.

[16] T. Zhou, S. Canu, P. Vera and S. Ruan, "Feature-enhanced generation and multi-modality fusion based deep neural network for brain tumor segmentation with missing MR modalities," *Neurocomputing*, vol. 466, pp. 102-112, 2021, doi: 10.1016/j.neucom.2021.09.032.

[17] Y. Liu, H. Ishibuchi, G.G. Yen, Y. Nojima and N. Masuyama, "Handling Imbalance Between Convergence and Diversity in the Decision Space in Evolutionary Multi-Modal Multi-Objective Optimization," *IEEE Trans. Evol. Comput.*, vol. 24, no. 3, pp. 551-565, 2020, doi: 10.1109/TEVC.2019.2938557.

[18] A.E.W. Johnson et al., "MIMIC-CXR, a de-identified publicly available database of chest radiographs with free-text reports," *Sci. Data*, vol. 6, pp. 317, 2019, doi: 10.1038/s41597-019-0322-0.

[19] A.E. Johnson et al., "MIMIC-CXR-JPG, a large publicly available database of labeled chest radiographs," *arXiv:1901.07042*, 2019.

[20] A. Johnson, L. Bulgarelli, T. Pollard, S. Horng, L.A. Celi and R. Mark, "Mimic-iv (version 1.0)," *PhysioNet*, 2020.

[21] G. Huang, Z. Liu, L. van der Maaten and K.Q. Weinberger, "Densely Connected Convolutional Networks," in *Proc. IEEE Conf. Comput. Vis. Pattern Recognit.*, IEEE, pp. 4700-4708, 2017.

[22] Y. Gu et al., "Domain-Specific Language Model Pretraining for Biomedical Natural Language Processing," *ACM Trans. Comput. Healthc.*, vol. 3, no. 1, pp. 1-23, 2022, doi: 10.1145/3458754.

[23] A. Vaswani et al., "Attention Is All You Need," in *Adv. Neural Inf. Process. Syst.*, pp. 5998-6008, 2017.

[24] Z. Liu, Y. Shen, V.B. Lakshminarasimhan, P.P. Liang, A. Zadeh and L. Morency, "Efficient Low-rank Multimodal Fusion with Modality-Specific Factors," 2018, *arXiv:1806.00064*.

[25] A. Zadeh, M. Chen, S. Poria, E. Cambria and L. Morency, "Tensor Fusion Network for Multimodal Sentiment Analysis," 2017, *arXiv:1707.07250*.

[26] J.H. Moon, H. Lee, W. Shin and E. Choi, "Multi-modal Understanding and Generation for Medical Images and Text via Vision-Language Pre-Training," *IEEE J. Biomed. Health Inform.*, vol. 26, no. 12, pp. 6070-6080, 2021.

[27] A. HaqueA. Milstein and L. Fei-Fei, "Illuminating the dark spaces of healthcare with ambient intelligence," *Nature*, vol. 585, no. 7824, pp. 193-202, 2020, doi: 10.1038/s41586-020-2669-y.

[28] Q. Suo, Z. Weida, M. Fenglong, Y. Ye, G. Jing and A. Zhang, "Metric Learning on Healthcare Data with Incomplete Modalities," in *Int. Joint Conf. Artif. Intell.*, pp. 3534-3540, 2019.

[29] C. Zhang et al., "M3Care: Learning with Missing Modalities in Multimodal Healthcare Data," in *Proceedings of the ACM international conference on knowledge discovery and data mining*, pp. 2418-2428, 2022.

[30] W. Ning et al., "Open resource of clinical data from patients with pneumonia for the prediction of COVID-19 outcomes via deep learning," *Nat. Biomed. Eng.*, vol. 4, no. 12, pp. 1197-1207, 2020, doi: 10.1038/s41551-020-00633-5.

[31] T. van Sonsbeek and M. Worring, "Towards Automated Diagnosis with Attentive Multi-modal Learning Using Electronic Health Records and Chest X-Rays," in *ML-CDS and CLIP (MICCAI)*, Springer, pp. 106-114, 2020.

[32] V. Singh, N.K. Verma, Z. Ul Islam and Y. Cui, "Feature Learning Using Stacked Autoencoder for Shared and Multimodal Fusion of Medical Images," in *Computational Intelligence: Theories Applications and Future Directions*, Springer, vol. 1, pp. 53-66, 2019.

[33] A. Dosovitskiy et al., "An Image is Worth 16x16 Words: Transformers for Image Recognition at Scale," in *Proc. Int. Conf. Learn. Represent.*, 2021.

[34] J. Devlin, M. Chang, K. Lee and T. Kristina, "BERT: Pre-training of Deep Bidirectional Transformers for Language Understanding," 2018, *arXiv:1810.04805*.

[35] W. KimB. Son and I. Kim, "ViLT: Vision-and-Language Transformer Without Convolution or Region Supervision," in *Int. Conf. Mach. Learn. (ICML)*, pp. 5583-5594, 2021.

[36] Z. Yu, J. Yu, Y. Cui, D. Tao and Q. Tian, "Deep Modular Co-Attention Networks for Visual Question Answering," in *Proc. IEEE Conf. Comput. Vis. Pattern Recognit.*, pp. 6281-6290, 2019.

[37] Y.H. Tsai, S. Bai, L.P. Pu, J.Z. Kolter, L.P. Morency and R. Salakhutdinov, "Multimodal Transformer for Unaligned Multimodal Language Sequences," in *Proc. Conf. Assoc. Comput. Linguistics*, vol. 2019, pp. 6558-6569, 2019.

[38] C. Li et al., "SemVLP: Vision-Language Pre-training by Aligning Semantics at Multiple Levels," *arXiv:2103.07829*, 2022.